\documentclass[twoside,twocolumn,english,prc,amsmath]{revtex4}
\usepackage[T1]{fontenc}
\usepackage{geometry}
\usepackage{amsmath}
\usepackage{graphicx}
\usepackage{amssymb}

\makeatletter
\def\fnum@table{\tablename~{\bf\thetable}}
\def\fnum@figure{\figurename~{\bf\thefigure}}
\def\tablename{\footnotesize{\bf Table}}
\def\figurename{\footnotesize{\bf Figure}}

\usepackage{dcolumn}

\usepackage{babel}
\makeatother
\begin{document}

\title{{\normalsize On the role of initial conditions and final state interactions
in ultrarelativistic heavy ion collisions}}

\author{{\normalsize K. Werner $^{(a)}$, T. Hirano $^{(b)}$, Iu. Karpenko
$^{(a,c)}$, T.$\,$Pierog$\,{}^{(d)}$, S. Porteboeuf $^{(a)}$,
M. Bleicher $^{(e)}$, S. Haussler $^{(e)}$}\\
{\normalsize ~}}

\address{$^{(a)}$ SUBATECH, University of Nantes -- IN2P3/CNRS-- EMN, Nantes,
France}

\address{$^{(b)}$ Department of Physics, University of Tokyo, Tokyo 113-0033,
Japan}

\address{$^{(c)}$ Bogolyubov Institute for Theoretical Physics, Kiev 143,
03680, Ukraine}

\address{$^{(d)}$ Forschungszentrum Karlsruhe, Institut fuer Kernphysik,
Karlsruhe, Germany}

\address{$^{(e)}$ Frankfurt Institute for Advanced Studies (FIAS), Johann
Wolfgang Goethe Universitaet, Frankfurt am Main, Germany}

\begin{abstract}
We investigate the rapidity dependence of the elliptical flow in heavy
ion collisions at 200 GeV (cms), by employing a three-dimensional
hydrodynamic evolution, based on different initial conditions, and
different freeze-out scenarios. It will be shown that the form of
pseudo-rapidity ($\eta$) dependence of the elliptical flow is almost
identical to space-time-rapidity ($\eta_{s}$) dependence of the initial
energy distribution, independent of the  freeze-out prescriptions.
\end{abstract}
\maketitle
A hydrodynamical treatment of ultrarelativistic heavy ion collisions
requires thermalized matter as an {}``initial condition'' at some
early time being of the order of a $\mathrm{fm/c}$. Such an initial
condition is difficult to access theoretically. There are also considerable
uncertainties concerning the evolution of the system, concerning transport
coefficients and the equation of state. Finally towards the end of
the evolutions, it seems more and more clear that the system will
not stay in thermal equilibrium, but interact nevertheless via hadronic
rescatterings, before freezing out.

It is therefore desirable to disentangle the different phases, try
to understand which kind of observables are sensitive to what feature
of the model description. In this paper we are going to investigate
the role of the initial condition, and we will in particular focus
on the rapidity dependence.

We will compare two options for initial conditions: a parameterization,
with parameters chosen in order to optimize final results \cite{tetsu},
referred to as {}``PAR'' thoughout this paper, and an initial condition
obtained from microscopic approach {}``EPOS'', based on the hypothesis
that thermalization happens very quickly and is achieved at some $\tau_{0}$.
For both options, we will perform three-dimensional hydrodynamic calculations,
using the same equation of state, see \cite{tetsu}. For either of
the two scenarios, we will investigate different freeze-out (FO) scenarios,
to be discussed later. This modular structure allows us to separate
initial and final state effects.

The PAR initial condition has been employed in many publications,
for details see \cite{tetsu}. In case of EPOS, the initial scatterings
lead to the formation of strings, which break into segments, which
are usually identified with hadrons. When it comes to heavy ion collisions,
the procedure is modified: one considers the situation at an early
proper time $\tau_{0}$, long before the hadrons are formed: one distinguishes
between string segments in dense areas (more than some critical density
$\rho_{0}$ segments per unit volume), from those in low density areas.
The high density areas are referred to as core, the low density areas
as corona \cite{core}. The corona is important for certain aspects,
not the ones looked at in this paper. So here we simply consider the
core part. In any case, it is important to note that initial conditions
from EPOS are based on strings, not on partons. Based on the four-momenta
of the string segments which constitute the core, we compute the energy
density $\varepsilon(\tau_{0},\vec{x})$ and the flow velocity $\vec{v}(\tau_{0},\vec{x})$. 

Having fixed the initial conditions, the system evolves according
the equations of ideal hydrodynamics, for details see \cite{tetsu}.
To have a flexible and mudular structure, we make first FO tables
(storing FO surface and flows) based on hydro calculations with PAR
and EPOS initial conditions (for given $T_{FO}$). We then generate
particles EbE from the core, using FO tables, based on the Cooper-Frye
prescription. 

In the following figures, {}``EPOS'' refers to the hydrodynamic
evolution based on EPOS initial conditions, {}``PAR'' refers the
parameterized initial conditions of \cite{tetsu}. Both calculation
use $\tau_{0}=0.6$fm/c, and the same equation of state.%
\begin{figure*}[htb]
\begin{center}\hspace*{-0.7cm}\includegraphics[%
  scale=0.72,
  angle=270]{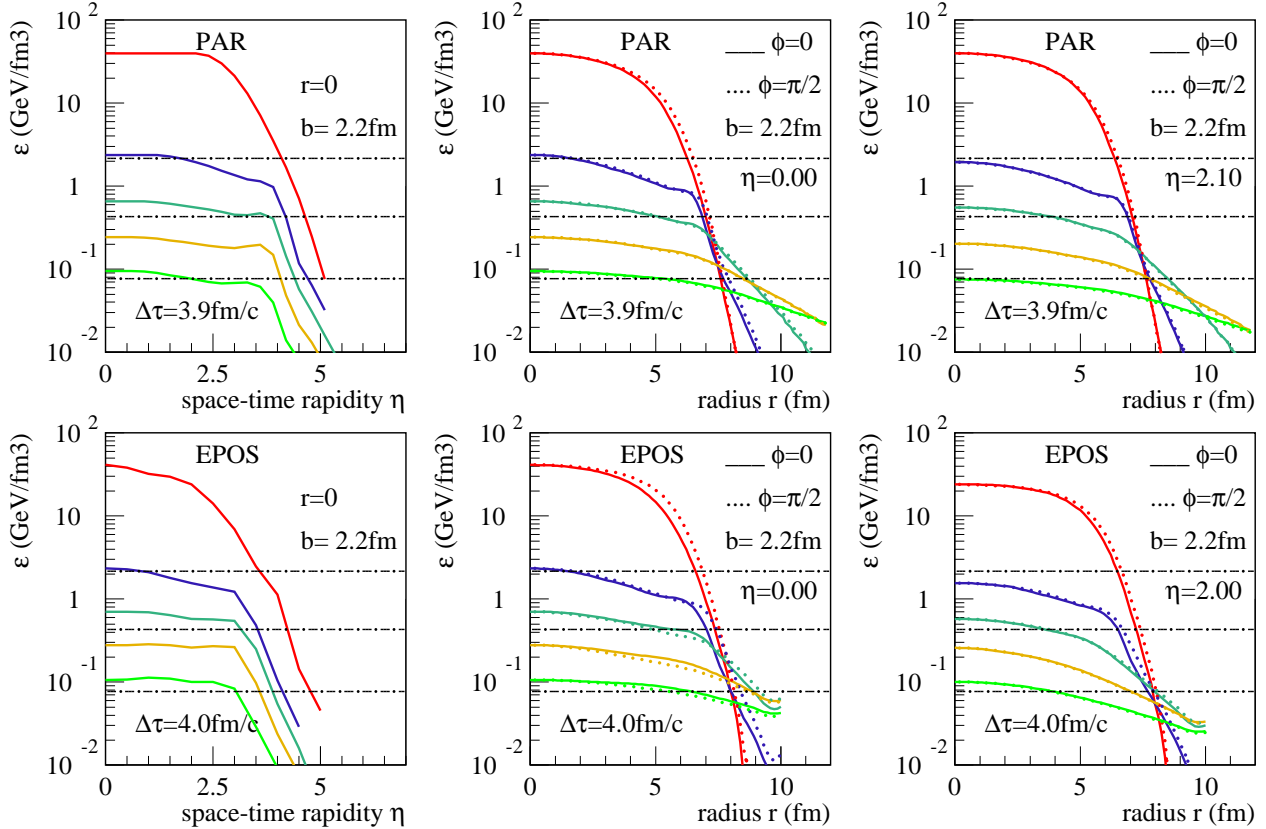}\end{center}

\caption{\label{cap:f1} Space-time evolutions of the energy density $\varepsilon$,
for a central Au+Au collision ($b=2.2\,$fm), for PAR (upper plots)
and EPOS (lower plots) initial conditions. In each plot, the different
curves refer to different times, from top to bottom: $\tau=\tau_{0}$,
$\tau=\tau_{0}+\Delta\tau$, $\tau=\tau_{0}+2\Delta\tau$, etc, with
$\Delta\tau=3.9\,$fm (4 fm) in case of PAR (EPOS). The left plots
show $\varepsilon$ as a function of the space-time rapidity, for
a transverse distance $r=0$, the middle plots show $\varepsilon$
as a function of the transverse distance $r$, for $\eta=0$, the
right ones for $\eta=2$. Full (dotted) curves refer to $\varphi=0$
($\varphi=\pi/2$). }
\end{figure*}
In fig. \ref{cap:f1}, we show the space-time evolutions of the energy
density $\varepsilon$, for a central Au+Au collision ($b=2.2\,$fm),
for PAR (upper plots) and EPOS (lower plots) initial conditions. In
each plot, the different curves refer to different times, from top
to bottom: $\tau=\tau_{0}$, $\tau=\tau_{0}+\Delta\tau$, $\tau=\tau_{0}+2\Delta\tau$,
etc, with $\Delta\tau=3.9\,$fm (4 fm) in case of PAR (EPOS). The
left plots show $\varepsilon$ as a function of the space-time rapidity
$\eta$, for a transverse distance $r=0$, the middle plots show $\varepsilon$
as a function of the transverse distance $r$, for $\eta=0$, the
right ones for $\eta=2$. Full (dotted) curves refer to $\varphi=0$
($\varphi=\pi/2$). The most striking difference between the PAR and
EPOS curves is the space-time rapidity dependence of the energy density,
at initial time $\tau_{0}$(upper curves on the left plots): Whereas
the PAR curve is flat up to $\eta=2$, the EPOS curve drops considerably. 

\begin{figure*}[htb]
\begin{center}\hspace*{-0.7cm}\includegraphics[%
  scale=0.72,
  angle=270]{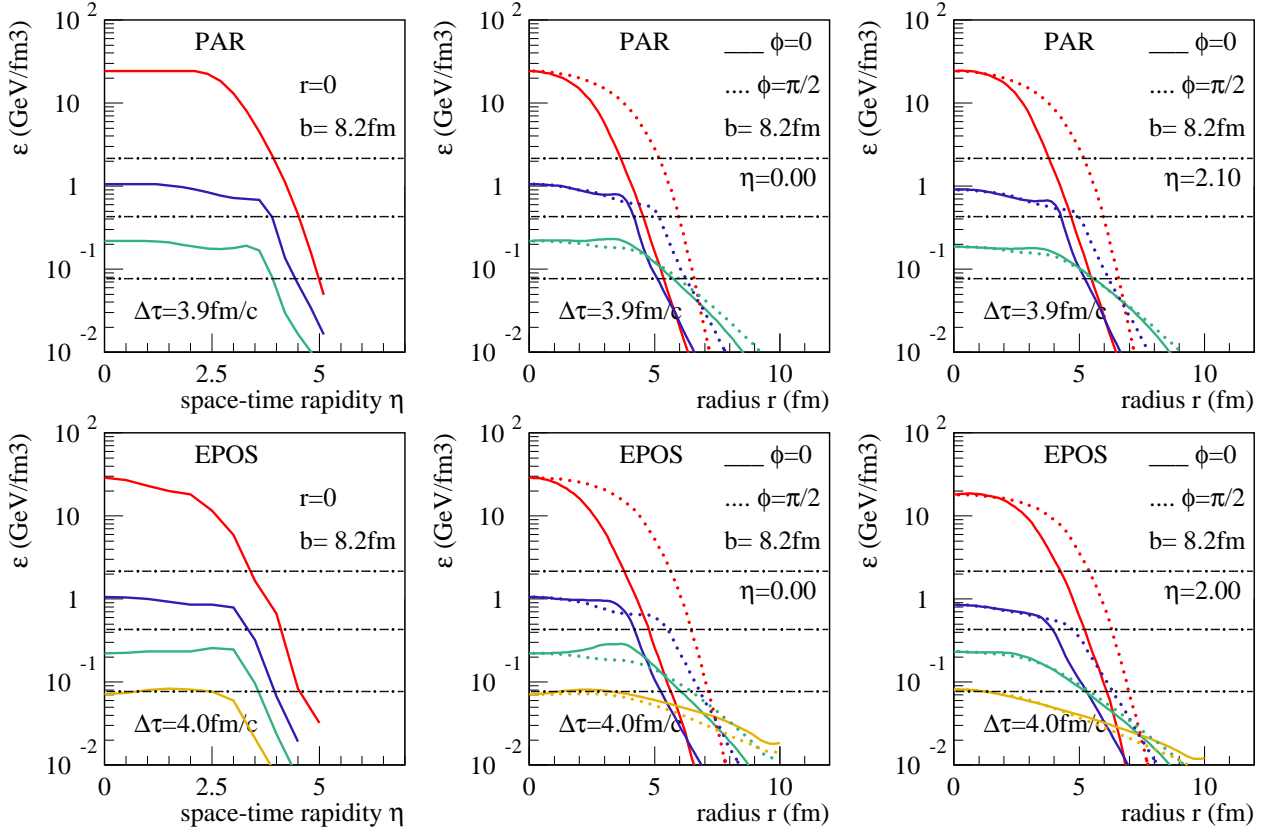}\end{center}

\caption{\label{cap:f2} Space-time evolutions of the energy density $\varepsilon$,
for a peripheral Au+Au collision ($b=8.2\,$fm), for PAR (upper plots)
and EPOS (lower plots) initial conditions. In each plot, the different
curves refer to different times, from top to bottom: $\tau=\tau_{0}$,
$\tau=\tau_{0}+\Delta\tau$, $\tau=\tau_{0}+2\Delta\tau$, etc, with
$\Delta\tau=3.9\,$fm (4 fm) in case of PAR (EPOS). The left plots
show $\varepsilon$ as a function of the space-time rapidity, for
a transverse distance $r=0$, the middle plots show $\varepsilon$
as a function of the transverse distance $r$, for $\eta=0$, the
right ones for $\eta=2$. Full (dotted) curves refer to $\varphi=0$
($\varphi=\pi/2$). }
\end{figure*}

In fig. \ref{cap:f2}, we show the space-time evolutions of the energy
density $\varepsilon$, for a peripheral Au+Au collision ($b=8.2\,$fm),
for PAR (upper plots) and EPOS (lower plots) initial conditions. Again,
the most striking difference is the stronger space-time rapidity dependence
of the initial distribution. In transverse direction the curves are
quite similar, however, the EPOS results have a bigger eccentricity,
as can be seen from the bigger difference between the $\varphi=0$
and the $\varphi=\pi$ curves. 

\begin{figure*}[htb]
\begin{center}\hspace*{-0.7cm}\includegraphics[%
  scale=0.72,
  angle=270]{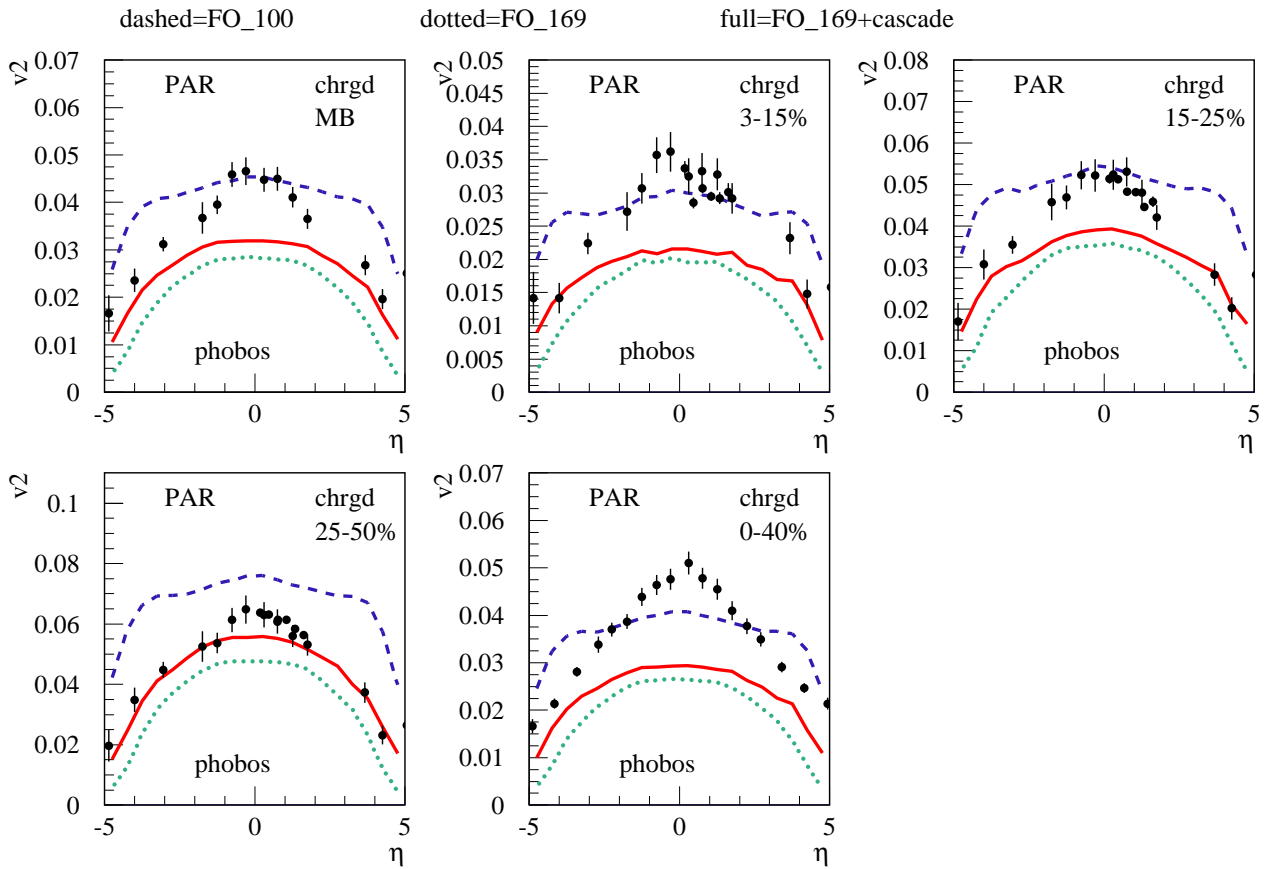}\end{center}

\caption{\label{cap:f3}The elliptical flow $v_{2}$ as a function of the
pseudorapidity $\eta$, for the PAR initial conditions, for minimum
bias (MB) events, as well as for different centrality classes: 3-15\%,
15-25\%, 25-50\%, and 0-40\% of the most central collisions, compared
to data \cite{phobos}. The three curves in each plot refere to different
freeze out prescriptions: FO\_100 (dashed), FO\_169 (dotted), FO\_169+cascade
(full).}
\end{figure*}

In the following, we will discuss results concerning the elliptical
flow $v_{2}$ as a function of the pseudorapidity $\eta$, for Au+Au
collisions at 200 GeV. For both PAR and EPOS, we show always three
curves: 

\begin{itemize}
\item a dashed one, representing a pure hydrodynamic evolution, with freeze
out 100 MeV (FO\_100);
\item a dotted curve, referring to a pure hydrodynamical evolutions, with
freeze out at 169 MeV (FO\_169);
\item a full curve, referring to a hydrodynamical evolution with freeze
out at 169 MeV, and subsequent hadronic cascade, using UrQMD (FO\_169+cascade).
\end{itemize}
We should keep in mind that the critical temperature is $T_{c}=170$
MeV . In fig.\ref{cap:f3}, we show first the $v_{2}$ results for
the PAR initial conditions, for minimum bias (MB) events, as well
as for different centrality classes: 3-15\%, 15-25\%, 25-50\%, and
0-40\% of the most central collisions, compared to data \cite{phobos}.
The data show for all centralities a more or less pronounced triangular
shape, with larger values for more peripheral collisions. The calculations
show as well a very similar shape for the different centrality classes,
for each of the three freeze out options. The FO\_100 option gives
the largest $v_{2}$ values, the shape is quite flat. The FO\_169
option (early freeze out) gives significantly smaller $v_{2}$ values,
and the distributions get narrower. Considering finally early freeze
out with subsequent hadronic cascade (FO\_169+cascade), leads again
to some increase of $v_{2}$, but it remains considerably lower than
the pure hydro results FO\_100.

The theoretical curves are identical to those shown in \cite{tetsu},
in case of the FO\_100 and FO\_169 options. Concerning the hadronic
rescattering, two different approaches have been employed: the UrQMD
model in this work, and JAM in ref. \cite{tetsu}. The results are
quite close, though not identical. The JAM curves show a slight dip
at midrapidity, which is absent when using UrQMD. It is difficult
to pin down the origin of these differences, but it is encouraging
that the differences are relatively small, and they may be considered
as the {}``systematic error'' of the theoretical treatment of this
part. 

\begin{figure*}[htb]
\begin{center}\hspace*{-0.7cm}\includegraphics[%
  scale=0.72,
  angle=270]{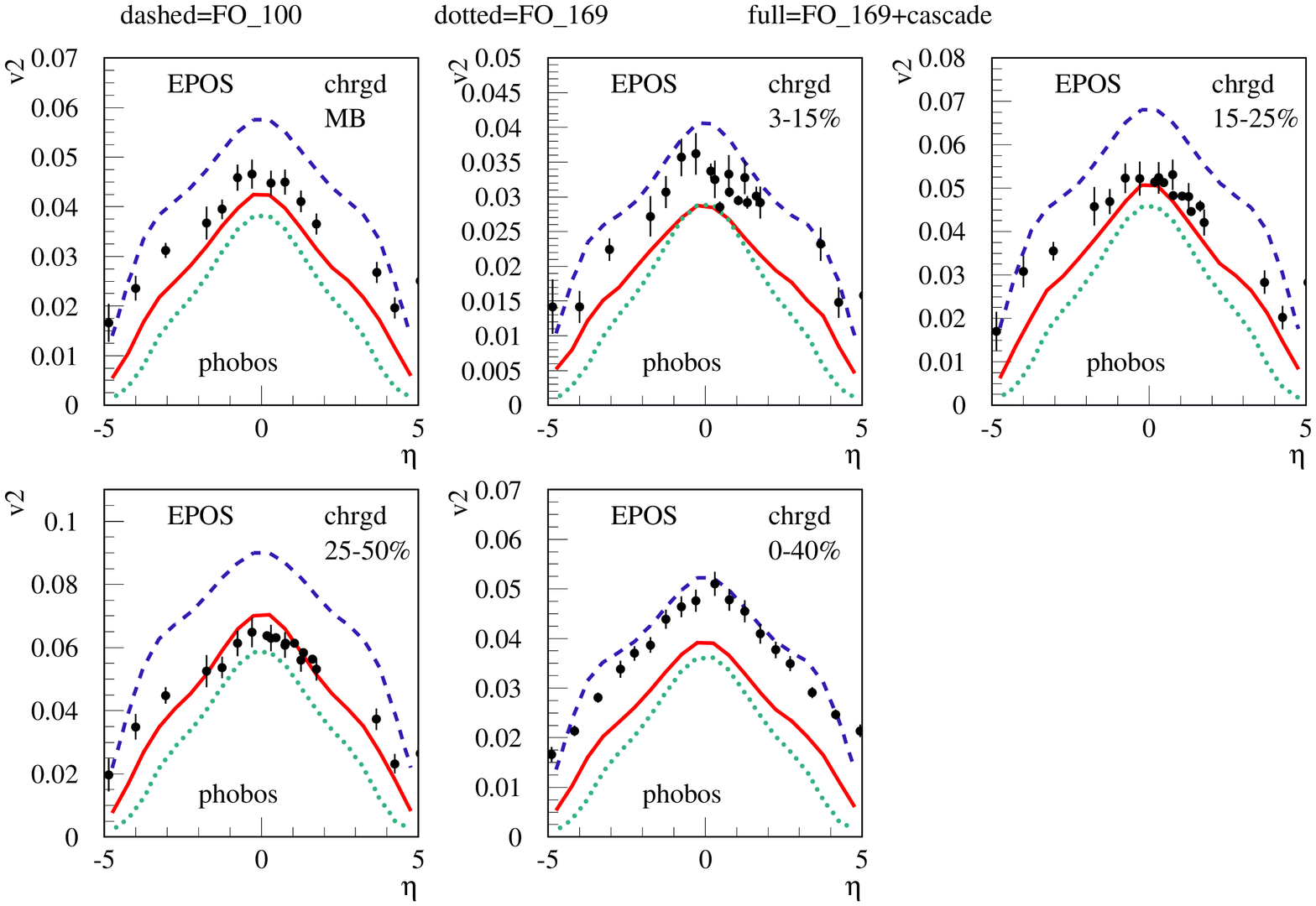}\end{center}

\caption{\label{cap:f4}The elliptical flow $v_{2}$ as a function of the
pseudorapidity $\eta$, for the EPOS initial conditions, for minimum
bias (MB) events, as well as for different centrality classes: 3-15\%,
15-25\%, 25-50\%, and 0-40\% of the most central collisions, compared
to data \cite{phobos}. The three curves in each plot refere to different
freeze out prescriptions: FO\_100 (dashed), FO\_169 (dotted), FO\_169+cascade
(full).}
\end{figure*}
In fig. \ref{cap:f4}, we show the corresponding $v_{2}$ results
for the EPOS initial conditions. Here, the shape is completely different
compared to the PAR initial conditions, it is more triangular. The
different freeze out options differ in magnitude (in the same fashion
as for the PAR results), but they all show a similar overall shape. 

How can we understand this large difference in the shape of the $\eta$
dependence of $v_{2}$, between PAR and EPOS initial conditions, which
is even independent of the freeze out prescriptions? From figs. \ref{cap:f1},\ref{cap:f2}
we know, that the space time evolution of the energy density in PAR
and EPOS are quite similar, apart of the fact that initially the space-time
rapidity $\eta_{s}$ dependence in PAR is flat, wheras in EPOS it
drops strongly with $\eta_{s}$. Using a linear scale, initial $\eta_{s}$
dependence is already almost triangular. So there is a strong correlation
between the $\eta_{s}$ width of the initial energy density and the
pseudorapidity distribution of $v_{2}$, which is compatible with
earlier findings \cite{tetsu2}.

The triangular-like shape of the initial $\eta_{s}$ distribution
in EPOS is, on the other hand, partly due to the fact that we use
strings as a basis of the calculation of the initial energy density.
Strings always strech over a certain range in $\eta_{s}$, with fluctuations
concerning the length of the string, but always covering $\eta_{s}=0$,
leading thus to a triangular-like shape.

\begin{figure*}[htb]
\includegraphics[%
  scale=0.7,
  angle=270]{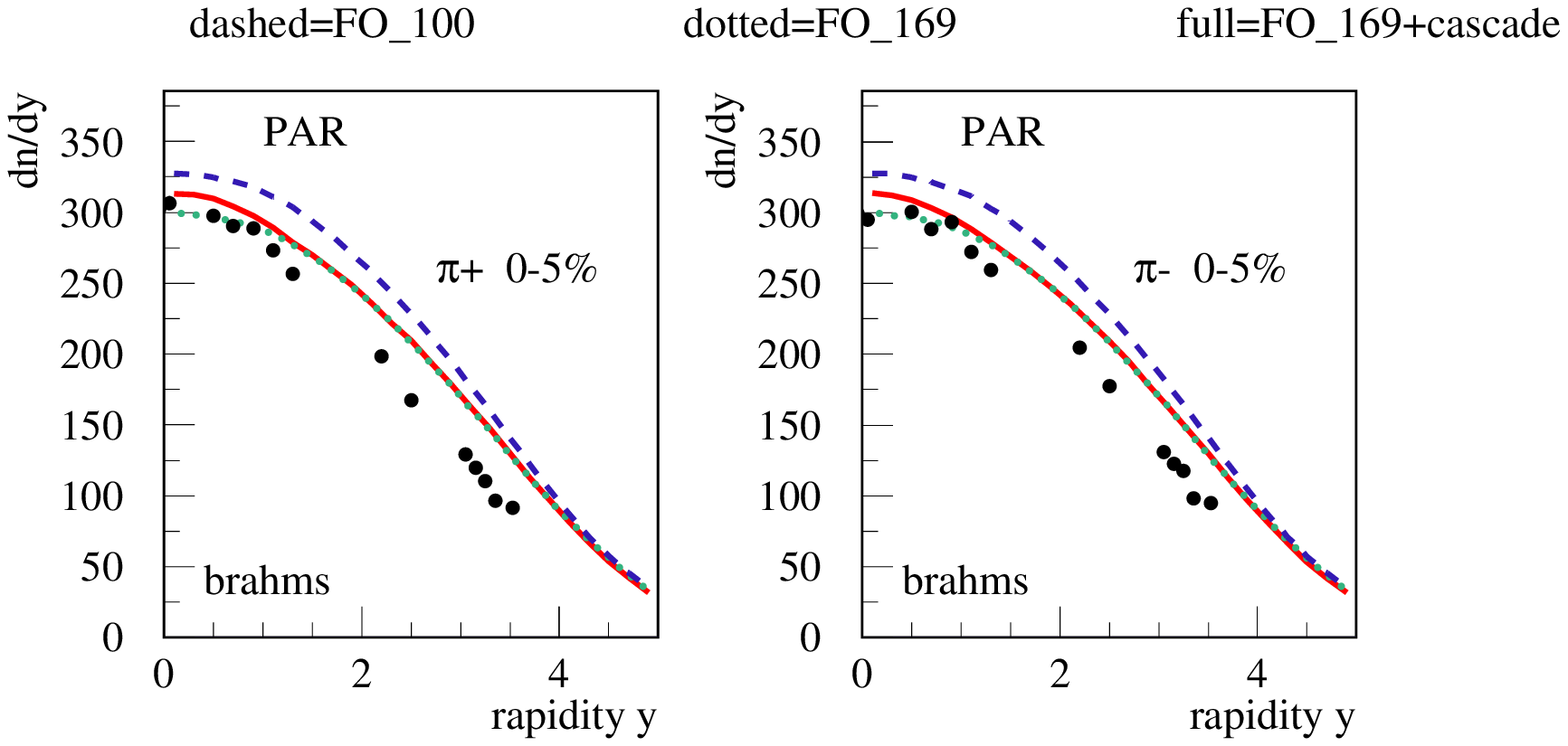}\\
\includegraphics[%
  scale=0.7,
  angle=270]{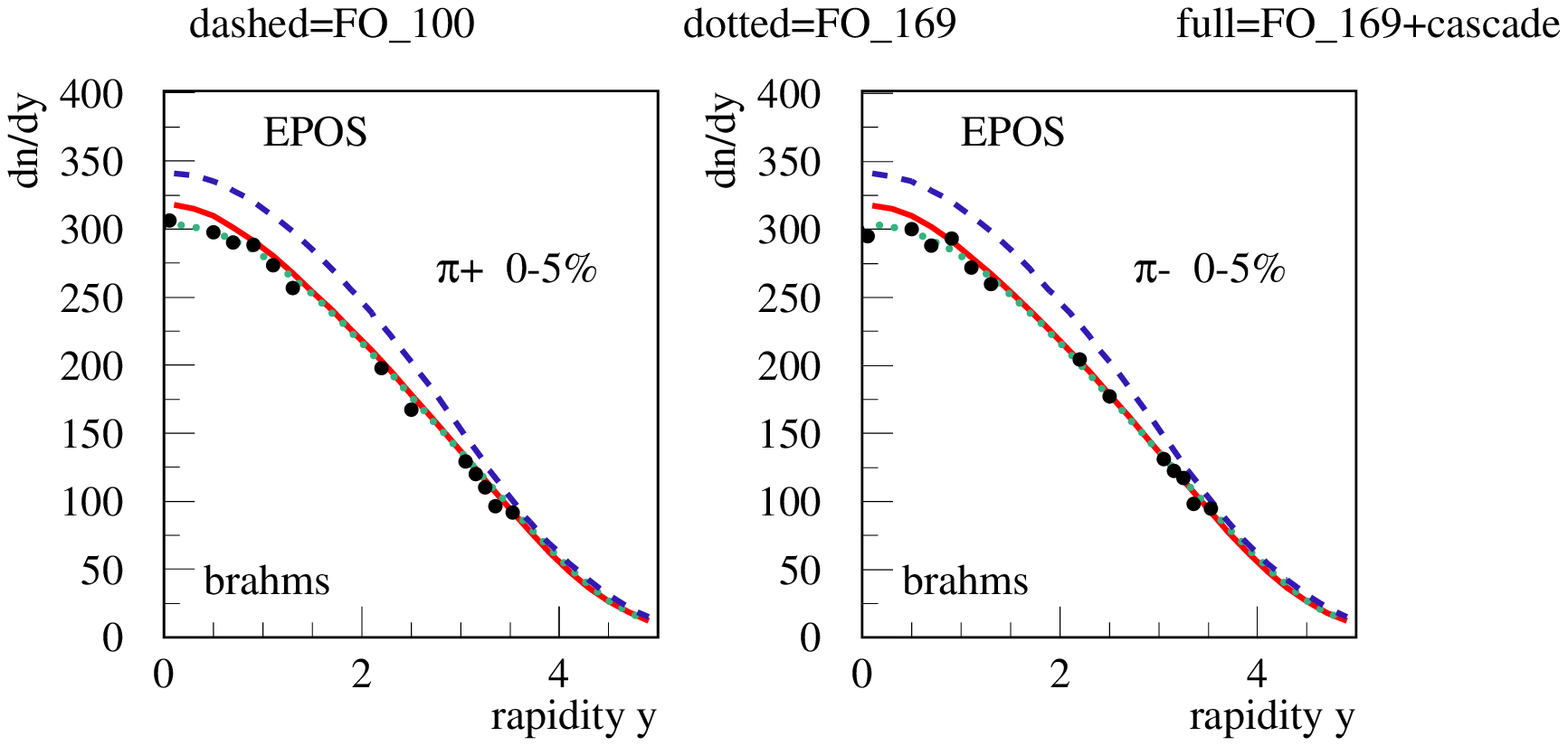}

\caption{Rapidity distributions of pions in central Au+Au collisions, for
the PAR initial conditions (upper panel), and EPOS initial conditions
(lower panel). The three curves in each plot refere to different freeze
out prescriptions: FO\_100 (dashed), FO\_169 (dotted), FO\_169+cascade
(full).\label{cap:Rapidity}}
\end{figure*}

We have seen significant differences in the rapidity dependence of
the elliptical flow, when choosing different initial conditions. Do
similar differences show up in simple rapidity spectra as well? In
fig. \ref{cap:Rapidity}, we show the rapidity distributions of pions
in central Au+Au collisions, for the PAR initial conditions (upper
panel), and EPOS initial conditions (lower panel). Again we consider
the three different freeze out scenarios, but they provide quite similar
results, although the late freeze out (FO\_100) gives slightly more
particles than the two other options, the latter ones being almost
identical. Comparing PAR and EPOS initial conditions, the former one
gives a broader distribution, as expected. But, the difference is
quite small, much smaller than the difference concerning the v2 rapidity
dependencies.

To summarize: we compared two options for initial conditions for 3D
hydrodynamical calculations of AuAu collisions: a parameterization,
with parameters chosen in order to optimize final results (PAR), and
an initial condition obtained from a microscopic approach (EPOS).
For both options, we performed the hydrodynamic calculations using
the same equation of state. For either of the two scenarios, we investigated
three different freeze-out options: a pure hydrodynamic evolution,
with freeze out 100 MeV, a pure hydrodynamical evolution, with freeze
out at 169 MeV, and a hydrodynamical evolution with freeze out at
169 MeV, with subsequent hadronic cascade. We found a fundamentally
different shape of the pseudorapidity dependence of $v_{2}$for the
two different initial conditions PAR and EPOS, independent of the
freeze out options. The characteristic triangular shape for the $v_{2}$
results for EPOS initial conditions is due to a triangular space-time
rapidity dependence of the initial energy density. The latter fact
is mainly due to the fact that strings are used to compute initial
conditions, not partons. It is interesting to note that rapidity spectra
are much less affected by the the choice of initial conditions than
the $v_{2}$ rapidity dependence.

\begin{acknowledgments}
T. H. is supported in part by the Grants-in-Aid of the Japanese Ministry
of Education, Culture, Sports, Science, and Technology (No.~19740130).
 The research has been carried out within the scope of the ERG (GDRE):
Heavy ions at ultra-relativistic energies - a European
Research Group comprising IN2P3/CNRS, Ecole des Mines de Nantes,
Universite de Nantes, Warsaw University of Technology, JINR Dubna, ITEP
Moscow and Bogolyubov Institute for Theoretical Physics NAS of Ukraine.
Iu. K. acknowledges the partial support of the Ministry for Education and
Science of Ukraine and Fundamental Research State Fund of Ukraine,
Agreement No F33/461-2009.

\end{acknowledgments}

\end{document}